\documentclass[aps,prb,twocolumn,superscriptaddress,preprintnumbers]{revtex4-2}
\usepackage[T1]{fontenc}
\usepackage{amssymb}
\usepackage{graphicx}
\usepackage{color}
\usepackage{caption}
\usepackage{ragged2e} 
\usepackage{hyperref}
\usepackage{amsthm}
\usepackage{amsmath}
\usepackage{braket}
\usepackage{appendix}
\usepackage{subcaption}
\usepackage{comment}
\usepackage{bbold}
\usepackage{multirow}
\usepackage{booktabs}
\usepackage{ulem}
\usepackage[ruled,vlined]{algorithm2e}
\usepackage[dvipsnames]{xcolor}
\usepackage{physics}
\usepackage{enumitem}
\usepackage{float}

\hypersetup{
    colorlinks = true,
    linkcolor = RoyalBlue,
    citecolor = RoyalBlue,
    urlcolor = RoyalBlue
}

\begin{document}

\title{Exploring the effect of basis rotation on the performance of Neural Quantum States}

\author{Sven Benjamin Kožić}
\email{skozic@irb.hr}
\affiliation{Institut Ru\dj er Bo\v{s}kovi\'c, Bijeni\v{c}ka cesta 54, 10000 Zagreb, Croatia}

\author{V. Zlatić}
\affiliation{Institut Ru\dj er Bo\v{s}kovi\'c, Bijeni\v{c}ka cesta 54, 10000 Zagreb, Croatia}
 
\author{F. Franchini}
\affiliation{Institut Ru\dj er Bo\v{s}kovi\'c, Bijeni\v{c}ka cesta 54, 10000 Zagreb, Croatia}

\author{S. M. Giampaolo}
\email{sgiampa@irb.hr}
\affiliation{Institut Ru\dj er Bo\v{s}kovi\'c, Bijeni\v{c}ka cesta 54, 10000 Zagreb, Croatia}

\date{\today}

\begin{abstract}

Neural Quantum States (NQS) are powerful variational representations of quantum many-body wavefunctions, yet their performance depends sensitively on the chosen basis. 
Using an exactly solvable one-dimensional Ising model, we show that local basis rotations leave the minimization landscape unchanged while relocating the exact ground state in parameter space.
This provides a controlled framework to disentangle representational limitations from optimization-induced trainability effects.
This geometric displacement, quantified through information-geometric measures, can steer optimization of shallow architectures toward saddle points and high-curvature regions. 
As a result, low energy errors may coexist with an incorrect wavefunction structure. 
By comparing energy and infidelity optimization within the same variational architectures, we show that optimization failure can persist even when the rotated target state remains representable. 
Our results identify a geometric mechanism contributing to basis dependence in NQS and motivate landscape-aware variational design.
\end{abstract}

\maketitle

\section{Introduction}

Understanding the ground-state properties of quantum many-body systems remains a central challenge in statistical physics and quantum information theory.
This difficulty originates from the exponential growth of the Hilbert space with system size, which makes exact representations of the state infeasible except for the smallest systems.
While a limited number of models admit analytic solutions~\cite{Baxter1982, Franchini2017}, most physically relevant systems do not.
Consequently, a wide range of approximate and variational approaches has been developed, including Quantum Monte Carlo methods~\cite{becca_sorella_2017}, tensor-network techniques~\cite{Orus2014}, and density functional theory~\cite{Kohn_1965}.

In recent years, neural-network-based variational approaches have emerged as a powerful and flexible framework for representing quantum many-body states.
Introduced by Carleo and Troyer~\cite{Carleo_2017}, Neural Quantum States (NQS) encode the amplitudes of a quantum state through a parametrized neural network, whose parameters are optimized variationally to approximate the ground state of a given Hamiltonian.
This framework combines variational Monte Carlo methods with the expressive power of neural-network parametrizations, enabling a flexible representation of many-body wavefunctions together with scalable optimization procedures.
Unlike tensor-network ans\"atze, neural networks can, in principle, represent states with volume-law entanglement~\cite{Deng_2017, Passetti_2023, Denis_2025}, and their training naturally benefits from parallelization and recent advances in machine-learning software and hardware~\cite{Vincentini_2022_01}.
Despite these advantages, the performance of NQS is known to depend strongly on the choice of basis used to represent the quantum state.
While basis dependence is an intrinsic feature of neural-network representations~\cite{Yang_2024}, the mechanisms through which a change of basis impacts variational optimization remain poorly understood.
Several recent works have investigated basis dependence in NQS~\cite{Pei_2021, Yang_2024, Cortes_2025}, as well as the role of optimization geometry and trainability limitations in variational neural-network ans\"atze~\cite{Chen_2024, Dash_2025, Moss_2025}. 
However, a controlled separation between changes in the physical properties of the target state and purely optimization-induced effects is still lacking.

With this work, we introduce a controlled and analytically transparent framework to isolate optimization-induced contributions to basis dependence from changes in the physical properties of the target state.
Our analysis employs the one-dimensional transverse-field Ising model with periodic boundary conditions, considering both ferromagnetic ($J=-1$) and antiferromagnetic ($J=+1$) couplings on chains with an odd number of spins.
In this setting, the distinction between ferromagnetic and antiferromagnetic regimes acquires a deeper structural significance.
In the ferromagnetic case, the system exhibits the familiar phenomenology of a symmetry-broken phase, characterized by a nearly degenerate low-energy doublet separated from the rest of the spectrum by a finite energy gap~\cite{Sachdev_2011}.
By contrast, an antiferromagnetic chain with periodic boundary conditions and an odd number of spins exhibits an irreducible \textit{topological frustration} (TF)~\cite{Maric2020_destroy,  Maric2020_induced, Maric2022_fate, Maric2022_nature}.
This frustration is induced solely by the boundary conditions and cannot be eliminated through any local rearrangement of interactions.
As a consequence, the low-energy sector forms a band of states whose energy gaps close polynomially with system size.
Previous works have shown that topological frustration gives rise to stable delocalized excitations and qualitative modifications of the ground-state structure, impacting not only dynamical properties~\cite{Torre2022} but also quantum resources~\cite{Giampaolo2019, Torre2024, Kozic2025, Odavic2023, Catalano2025}.
Furthermore, the robustness and genuinely global character of topological frustration render these effects experimentally accessible in state-of-the-art platforms, including Rydberg-atom systems~\cite{catalano2025_experiment}, thereby opening the way to concrete realizations of technological devices~\cite{Catalano2024}.
The contrast between the ferromagnetic and frustrated regimes therefore provides a controlled setting to investigate how different low-energy spectral structures and basis choices influence the trainability of variational quantum ans\"atze.

We employ shallow neural quantum state architectures, namely Restricted Boltzmann Machines (RBM) and small feedforward networks, trained via stochastic reconfiguration (quantum natural gradient)~\cite{Stokes_2020}. 
To isolate optimization-induced geometric effects, we apply a site-resolved local basis rotation, parameterized by an angle $\phi$, to the exact ground state.
This transformation preserves the physical spectrum and entanglement structure of the system while relocating the target state relative to the variational parametrization.
Our results show that shallow NQS architectures trained with quantum natural gradient are particularly sensitive to this relocation.
Depending on the rotation angle, optimization trajectories can be diverted toward saddle-point regions or poorly conditioned regions of the variational landscape.
In these regimes, the variational optimization may converge to states with small relative energy errors while failing to faithfully reproduce the correct wavefunction structure.
To diagnose this mismatch, we complement energy-based metrics with fidelity and coefficient Shannon entropy (quantum coherence)~\cite{Kozic_2025}, which provide a more complete characterization of the learned state.

Beyond quantum many-body physics, our analysis naturally connects to broader questions in modern machine learning concerning optimization landscapes and information geometry~\cite{Belkin_2019,Roberts_2022,Moss_2025}. 
Recent studies suggest that the trainability of NQS is strongly influenced not only by representational expressivity, but also by tthe geometry induced by the variational objective and the optimizer employed.
In particular, improvements in expressive capacity do not necessarily translate into efficient optimization, especially in the presence of ill-conditioned or weak-curvature regions of the variational landscape~\cite{Chen_2024,Giuliani_2023,Dash_2025}.

The paper is organized as follows.
In Sec.~II we introduce the basis-rotation framework together with the NQS architectures, cost functions, and optimization methods employed throughout this work.
In Sec.~III we develop the geometric perspective underlying our analysis and illustrate it using a toy model that captures the optimization mechanisms observed in the main numerical results.
Sections~IV and V present the numerical results and their implications for variational optimization in shallow NQS architectures.
Finally, in Sec.~VI we discuss the broader implications of our findings in the context of variational quantum algorithms and relations between trainability and geometry.

\section{Methods}\label{sec:Methods}

\subsection{Rotated Ising framework}
\label{sec:Basis Rotation}

To isolate how basis choice affects NQS optimization without modifying the underlying spectral properties of the system, we study the transverse-field Ising model on a ring:
\begin{equation}\label{eq:Hamiltonian}
    H = J \sum_{i=1}^{N} \sigma^{z}_{i}\sigma^{z}_{i+1} + h \sum_{i=1}^{N} \sigma^{x}_{i},
\end{equation}
where $\sigma^{x}$ and $\sigma^{z}$ are the usual Pauli matrices and periodic boundary conditions identify site $N+1\equiv 1$.
We then apply a local rotation around the $y$ axis at each site.
The single-spin rotation operator and the corresponding global rotation acting on the full chain are defined, respectively, as
\begin{equation}
    R^{y}_i(\phi):=e^{\imath \phi \sigma^{y}_i}, \qquad U^{y}(\phi):=\bigotimes_{i=1}^{N}R_i^{y}(\phi)
\end{equation}
Under this transformation, the Hamiltonian is mapped to $H(\phi)=U_{\phi}H U_{\phi}^{\dagger}$,
which preserves the energy spectrum while rotating the eigenstates in Hilbert space.
In particular, if $\lvert \psi_{0}\rangle$ is the ground state in the computational ($\sigma^{z}$) basis, then $\lvert \psi_{\phi}\rangle = U^{y}(\phi)\lvert \psi_{0}\rangle$ is the corresponding ground state of the rotated Hamiltonian $H(\phi)$.

\begin{figure*}[t]
\centering
\includegraphics[width=0.9\linewidth]{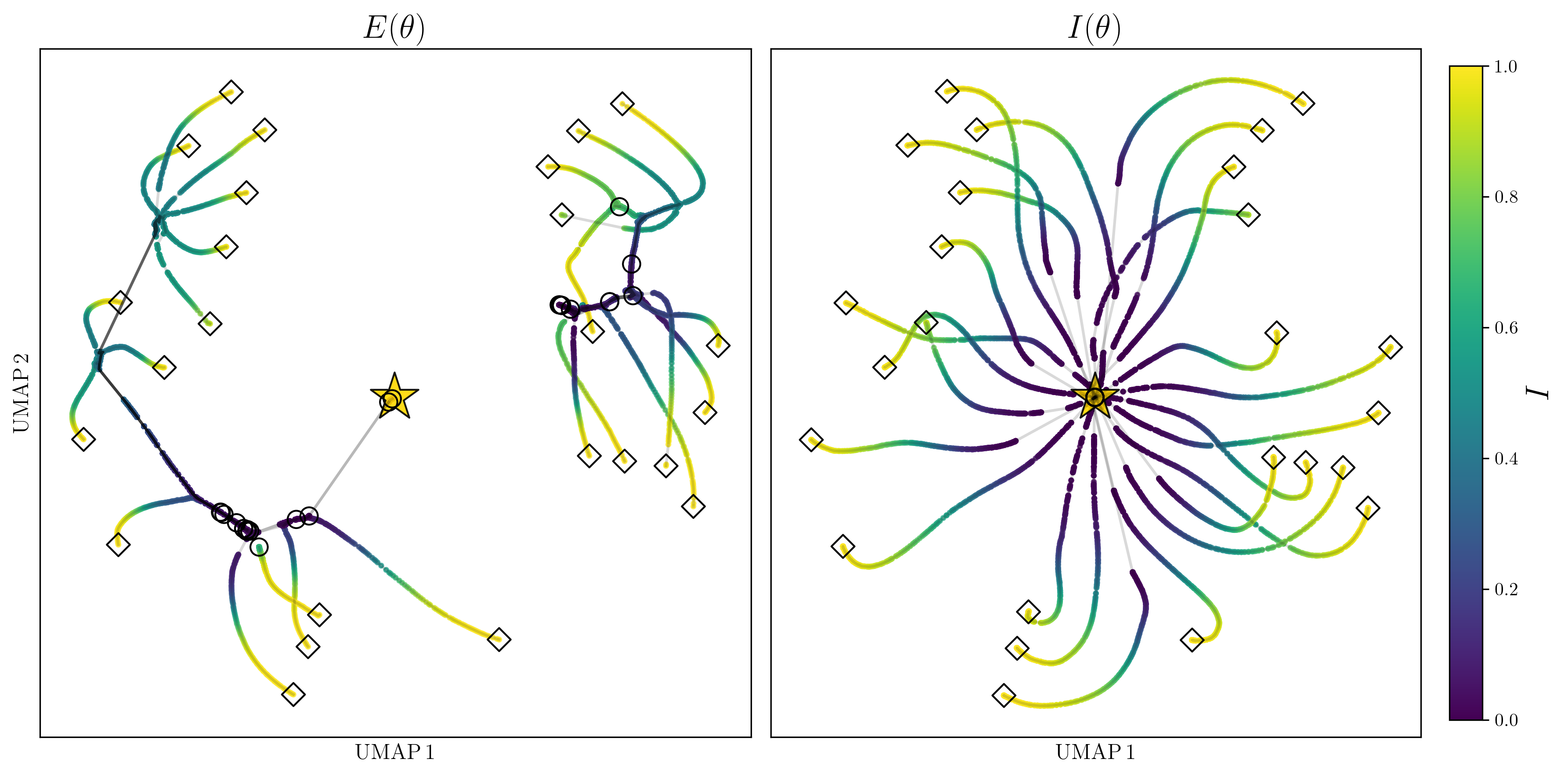}
\caption{\justifying
UMAP embedding of statevector trajectories for a complex-valued RBM NQS with $\alpha=2$ trained on the transverse-field Ising model in Eq.~\eqref{eq:Hamiltonian} with periodic boundary conditions, for $N=5$ and couplings $(J,h)=(-1.0,-0.5)$. 
We compare optimization using the energy cost function in Eq.~\eqref{eq:energy} (left panel) and the infidelity cost function in Eq.~\eqref{eq:infidelity} (right panel). 
For each case, $25$ independently initialized trajectories are optimized using stochastic reconfiguration for $5\times10^{3}$ steps; paired trajectories share the same random initialization ($\sigma=0.5$). 
Starting and final points are marked by circles $\circ$ and diamonds $\diamondsuit$, respectively, while the exact ground state is indicated by the golden star $\star$. 
Points along each trajectory are colored according to the infidelity $I\in[0,1]$ with respect to the exact ground state. 
Energy optimization exhibits clustering and trapping of trajectories, whereas infidelity optimization produces a substantially smoother convergence toward the target state.}
\label{fig:umap_trajectories}
\end{figure*}

\subsection{Variational ans\"atze and optimization geometry}

Throughout this work we consider two classes of neural-network variational states, both providing flexible variational parametrizations for correlated quantum many-body systems~\cite{Deng_2017,Passetti_2023}.
In both cases, denoting by $\theta$ the set of complex-valued variational parameters, the variational wavefunction is written as
\begin{equation}
  \lvert \psi_{\theta}\rangle 
  \;=\; \sum_{s} \psi_{\theta}(s)\,\lvert s\rangle,
  \quad
  \psi_{\theta}(s) = \exp\bigl[f_{\theta}(s)\bigr].
\end{equation}
The logarithmic parametrization naturally defines the score functions
\begin{equation}
  O_{k}(s) \;=\; \partial_{k}\ln\psi_{\theta}(s),
  \quad
  \partial_{k}\psi_{\theta}(s) = O_{k}(s)\,\psi_{\theta}(s),
\end{equation}
which are central to gradient-based optimization in high-dimensional parameter spaces~\cite{Vicentini_2022}.

The first architecture we consider is a Restricted Boltzmann Machine (RBM) with $N$ visible units and $M$ hidden units. 
In this case the log-amplitude reads
\begin{equation}
  f_{\theta}(s) 
  = \sum_{i=1}^{N} a_{i}\,s_{i} 
  +
  \sum_{j=1}^{M} \ln\Bigl[1 + \exp\bigl(\sum_{i=1}^{N} W_{ij}s_{i} + b_{j}\bigr)\Bigr],
\end{equation}
where $\theta=\{a_{i},b_{j},W_{ij}\}$ and the hidden-unit ratio $\alpha_{\mathrm{RBM}}=M/N$ controls the expressive capacity of the ansatz.

Alongside RBMs, we also consider shallow fully connected feedforward neural networks (FCNNs), where the depth $L$ and hidden-layer widths $\{n_{\ell}\}$ determine the expressive capacity of the variational state:
\begin{align}
  h^{(\ell)}_{j}(y)
  &= \sum_{i=1}^{n_{\ell-1}} W^{(\ell)}_{ij}\,y_i + b^{(\ell)}_j \nonumber\\ 
  f_\theta(s) &= \bigl[h^{(L)}\circ \cdots \circ (\varphi\circ h^{(1)})\bigr](s)
\end{align}
where $\varphi$ is a nonlinear activation function (e.g.\ $\tanh$), symbol $\circ$ denotes function composition, and $\theta=\{W^{(\ell)}_{ij},\,b^{(\ell)}_j\}_{\ell=1}^{L}$.

To disentangle optimization effects from representational limitations, we consider two complementary objective functions for the variational optimization process, namely the variational energy
\begin{equation}\label{eq:energy}
E(\theta)=\frac{\langle \psi_\theta|H|\psi_\theta\rangle}{\langle\psi_\theta|\psi_\theta\rangle},
\end{equation}
and, whenever the exact target state $|\psi_t\rangle$ is available, the infidelity
\begin{equation}\label{eq:infidelity}
I(\theta)=1-\frac{|\langle\psi_\theta|\psi_t\rangle|^2}{\langle\psi_\theta|\psi_\theta\rangle\,\langle\psi_t|\psi_t\rangle}.
\end{equation}

Optimization is performed using stochastic reconfiguration (quantum natural gradient), where the geometry of the variational manifold is encoded in the quantum geometric tensor~\cite{Stokes_2020}
\begin{equation}
G_{ij}(\theta)
= \frac{\langle \partial_{i}\psi_{\theta}|\partial_{j}\psi_{\theta}\rangle}{\langle \psi_{\theta}|\psi_{\theta}\rangle}
-\frac{\langle \partial_{i}\psi_{\theta}|\psi_{\theta}\rangle\,\langle \psi_{\theta}|\partial_{j}\psi_{\theta}\rangle}{\langle \psi_{\theta}|\psi_{\theta}\rangle^{2}}.
\end{equation}
In this framework, an infinitesimal variation of the parameters induces a Fubini--Study distance \begin{equation}
d^2_{\mathrm{FS}}\simeq \delta\theta^\dagger G(\theta)\delta\theta,
\end{equation} 
which determines the natural metric structure explored during optimization. 
The resulting update rule takes the form
\begin{equation}
    \theta \leftarrow \theta - \eta\,[G(\theta)+\epsilon I]^{-1}\nabla_\theta \mathcal{L},
\end{equation}
where $\mathcal{L}\in\{E,I\}$ denotes the objective function, $\eta$ the learning rate, and $\epsilon$ a Tikhonov regularization parameter.

To visualize how optimization trajectories depend on the choice of objective function, we employ Uniform Manifold Approximation and Projection (UMAP)~\cite{McInnes_2018,umap-learn} to embed statevector trajectories into two dimensions.
Figure~\ref{fig:umap_trajectories} shows that energy optimization produces clustered trajectories and slow convergence toward the target state, whereas infidelity optimization yields substantially smoother optimization paths.
Although expected, this distinction anticipates the geometric mechanisms underlying the basis-dependent optimization effects discussed in the following sections.

\begin{figure*}[t]
\centering
\includegraphics[width=1.0\linewidth]{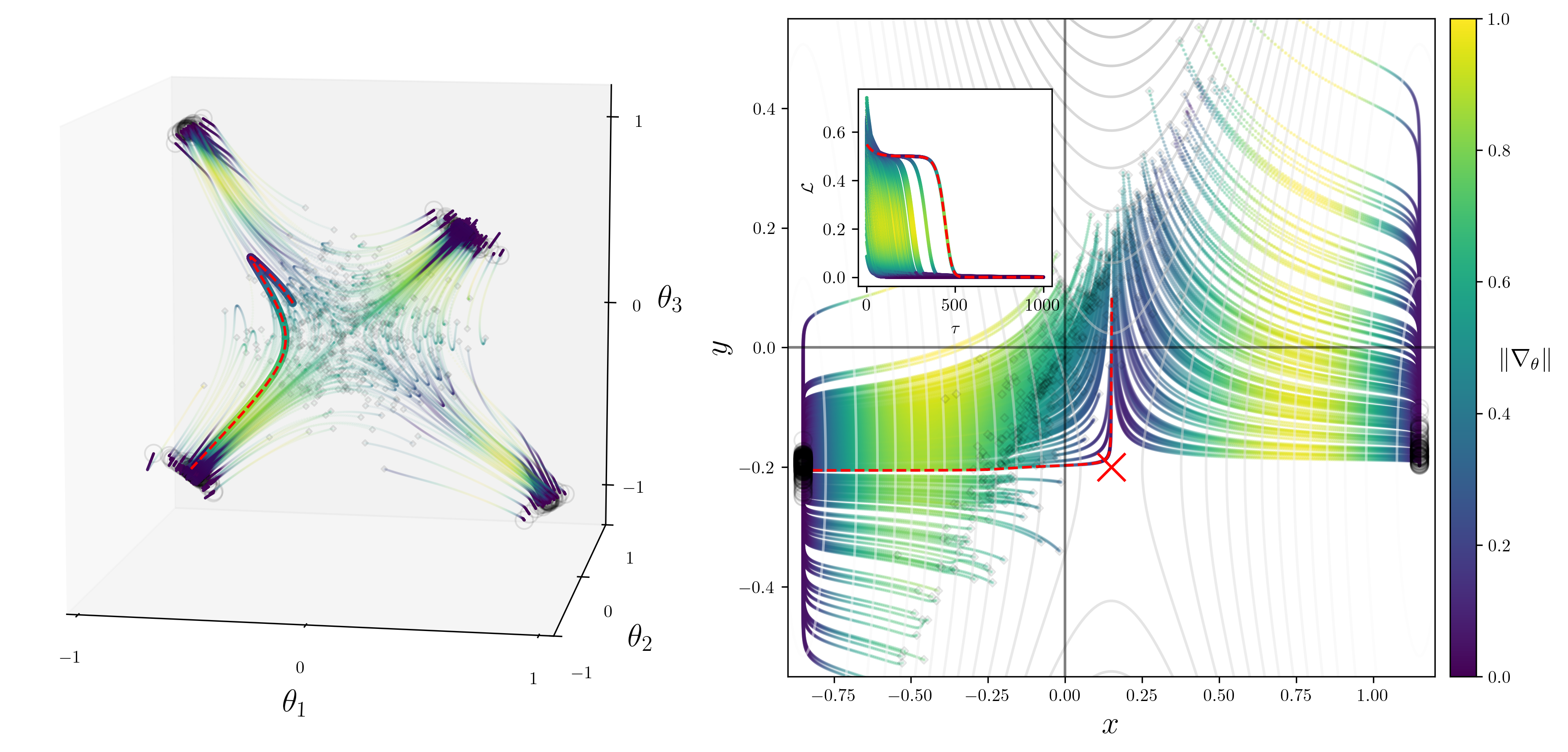}
\caption{\justifying 
Toy example illustrating how a saddle in the loss landscape can trap optimization trajectories and induce slow dynamics in parameter space.
We consider a two–dimensional objective function $f(x,y)$ with a double-well potential and a saddle point at $(u_s,v_s) = (0.15, -0.2)$.
A small neural network maps two fixed scalar inputs to a two-dimensional output space $(u,v)$ and is optimized using natural-gradient updates.
Starting from multiple random initializations, we track both the trajectories in output space and the corresponding evolution in parameter space.
Left panel: optimization trajectories in parameter space, shown as a dense cloud with color indicating the gradient norm (darker points correspond to smaller norms), together with start/end markers and one representative trajectory (red) that remains close to the saddle for an extended time.
Right panel: the same set of runs projected into the output space and overlaid on contour lines of $f(x,y)$; the saddle location is marked by a cross and the selected "stuck" trajectory is highlighted in red.
The inset reports the loss evolution during optimization.
The figure shows how saddle regions in the output landscape correspond to extended slow regions in parameter space.}
\label{fig:toy_example}
\end{figure*}

\section{Optimization geometry and variational dynamics}
\label{sec:considerations}

A central difficulty in comparing different variational approaches, including neural networks, tensor networks, and variational quantum circuits, lies in distinguishing limitations of the optimization procedure from limitations of the variational ansatz itself. 
Indeed, models with comparable expressive capacity may display substantially different convergence properties, suggesting that optimization geometry plays a fundamental role in determining trainability.

To discuss the performance of neural network learning in general, the most convenient setup arises naturally from differential geometry, which involves treating the network as a smooth map.~\cite{Amari_2000}
\begin{equation}\label{eq:smooth_map}
    f:\Theta\,\to\,\mathcal{Y}
\end{equation}
Let us for simplicity consider the space of real parameters $\Theta\subset\mathbb{R}^{p}$ (with coordinates $\theta$) and $\mathcal{Y}\subset\mathbb{R}^{D}$.
An infinitesimal change $v\in T_{\theta}\Theta$ is carried forward by the differential (pushforward)~\cite{Lee_2003}
\begin{equation}
    df_{\theta}(v)=\left.\frac{d}{dt}f(\theta + tv)\right|_{t=0}
\end{equation}
which is computed in automatic differentiation via a Jacobian--vector product  (JVP)~\cite{Baydin_2017,Jax_2018}. 
Conversely, a cotangent vector $w \in T^{*}_{f(\theta)} \mathcal{Y}$ is pulled back to a covector on $\Theta$ by the pullback map 
\begin{equation}
df_{\theta}^*(w)=J(\theta)^\top w,
\end{equation}
where $J(\theta)=\frac{\partial f(\theta)} {\partial\theta}$ is the Jacobian.
If we equip $\mathcal{Y}$ with a Riemannian metric $G_y$ then its pullback through $f$ induces a metric $g_{\theta}$ on $\Theta$:
\begin{equation}
g_{\theta}(v,v)=\langle df_{\theta}(v),df_{\theta}(v)\rangle_{G_{y}} =v^\top J(\theta)^\top G_{y}J(\theta)v
\end{equation}
Now let $\ell:\mathcal{Y}\to\mathbb{R}$ be a scalar loss defined on outputs, and define the generic objective function $\mathcal{L}(\theta)=\ell\bigl(f(\theta)\bigr)$.
In Euclidean coordinates on $\mathcal{Y}$, the gradient of $\mathcal{L}$ with respect to parameters is the pullback of the output differential, yielding the usual chain rule
\begin{equation}
   \nabla_{\!\theta}\mathcal{L}(\theta)=J(\theta)^\top \nabla_{y}\ell\bigl(f(\theta)\bigr) 
\end{equation}
The natural gradient method uses the metric $g_{\theta}$ (and, in statistical models, often the Fisher information) to precondition this ordinary gradient:
\begin{equation}
\widetilde{\nabla}\mathcal{L}(\theta)=g_{\theta}^{-1}\,\nabla_{\!\theta}\mathcal{L}(\theta),
\end{equation}
\begin{figure*}
  \centering
  \includegraphics[width=0.7\textwidth]{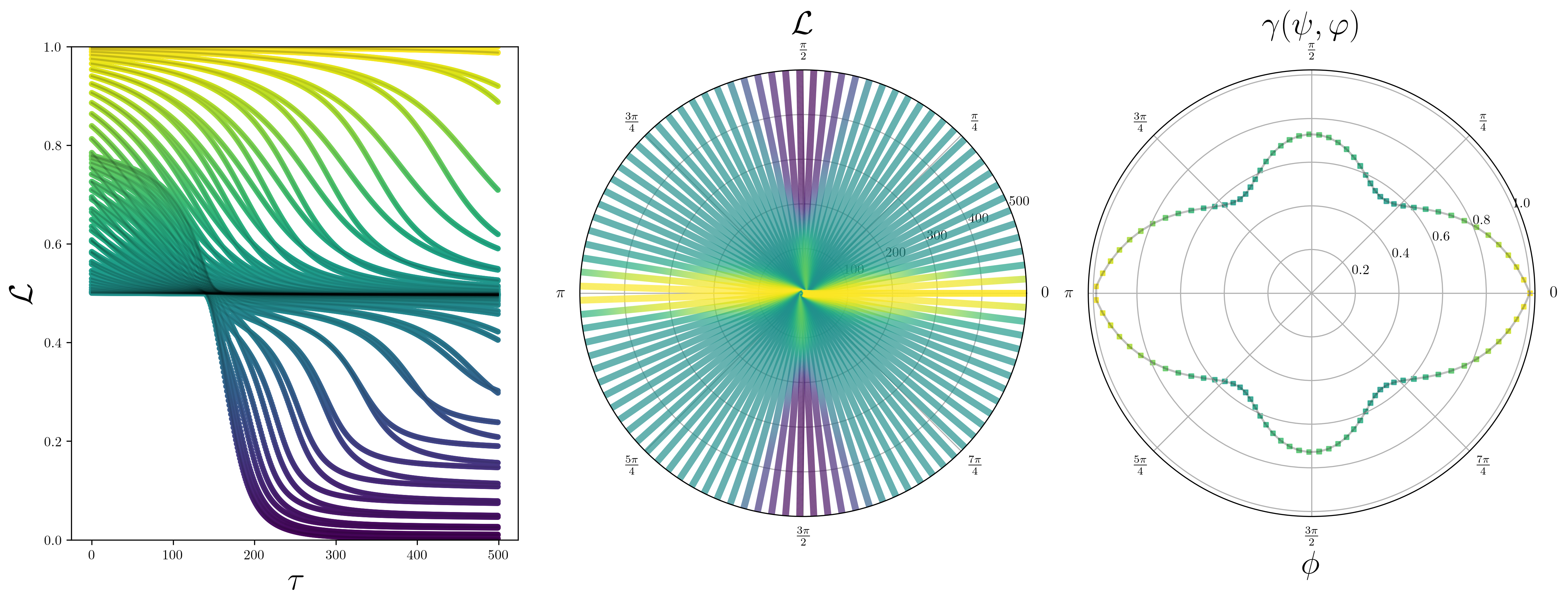}\par\bigskip
  \includegraphics[width=0.7\textwidth]{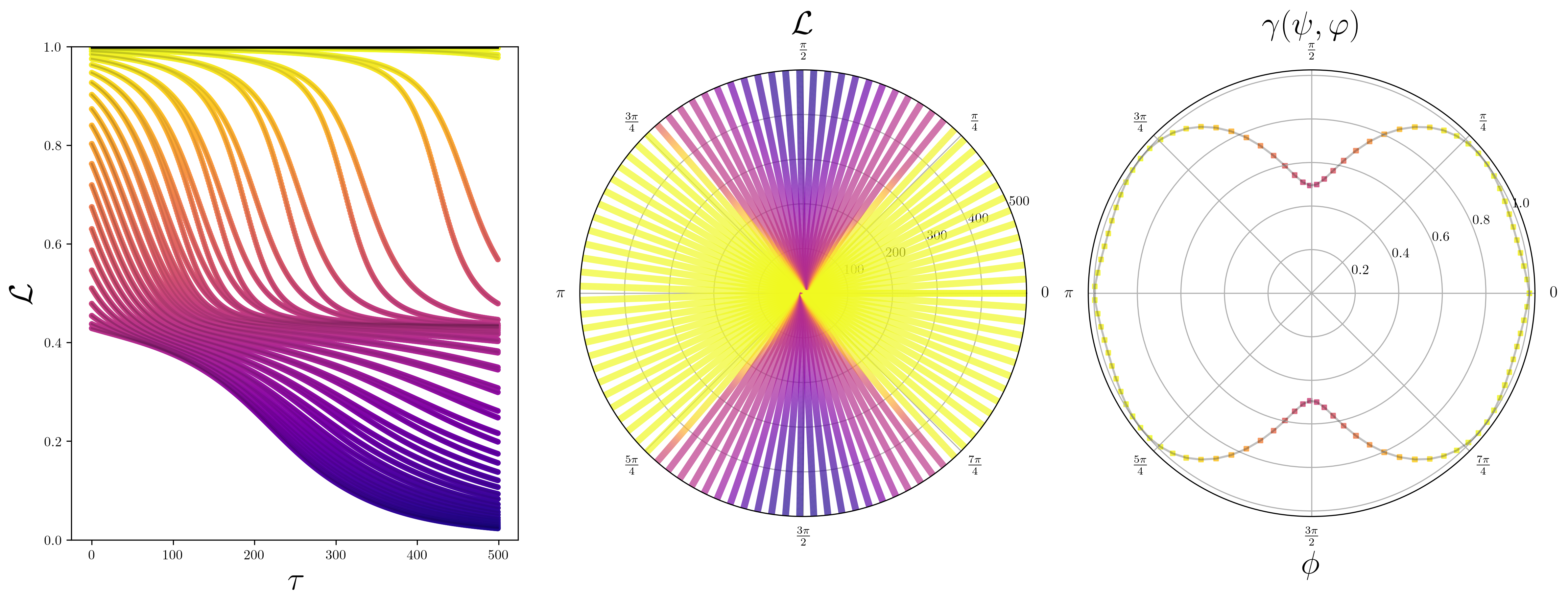}
\captionof{figure}{\justifying{\bf Infidelity minimization:} Comparison between the Fubini--Study distance $\gamma$ and the infidelity cost function $\mathcal{L}$ for NQS trained via infidelity minimization toward the rotated ground state $\lvert\varphi(\phi)\rangle$, for $N=5$, $h=0.5$, with $J=-1$ (top panels) and $J=+1$ (bottom panels), as a function of the rotation angle $\phi$. 
The left column shows the infidelity as a function of the optimization step $\tau$, while the middle column displays the same data in polar coordinates to emphasize the dependence on $\phi$. 
The right column reports the corresponding Fubini--Study distance $\gamma$ between the initial and target states. 
Optimization slows down for rotation angles where the target state becomes maximally separated from the initialization, although the convergence dynamics also depend on the local curvature of the optimization landscape.}
  \label{fig:FS_comparison}
\end{figure*}
In this perspective, optimization difficulties emerge when the pullback geometry induced by the variational ansatz generates poorly conditioned directions or extended low-curvature regions in parameter space, even when the target state itself remains representable.

This update follows the direction of steepest descent measured with respect to the geometry defined by $g_{\theta}$, and it is invariant under smooth changes of coordinates in parameter space~\cite{Amari_1998,Martens_2020}.
From this perspective, optimization generates a trajectory $\{\theta_t\}\subset\Theta$ in parameter space and, simultaneously, an induced trajectory $\{y_t\}=\{f(\theta_t)\}\subset\mathcal{Y}$ in output space. 
Pushforward describes how parameter perturbations move outputs, while pullbacks describe how loss information defined on $\mathcal{Y}$ is transferred back to $\Theta$ to define update directions.~\cite{Jax_2018, Amari_1998}.

To illustrate the dual trajectory, in Fig.~\ref{fig:toy_example} we use a two–dimensional output space ($N=1$) with coordinates $(x,y)$ and define the training objective as the scalar function 
\begin{eqnarray}
    f(x,y)\!=\!\tfrac{1}{2}\Bigl(\![(x\!-\!u_s)^2\!-\!1]^2\![1\!+\!(y\!-\!v_s)^2]\!-\!v_s\!(y\!-\!v_s)^2\!\Bigr)\!,
\end{eqnarray}
with $u_s=0.15$ and $v_s=-0.20$.
The minima of the function $f(x,y)$ are at $(x,y)=(u_s\pm1,v_s)$ with $f=0$ and the saddle is at $(x,y)=(u_s,v_s)$ with $f=0.5$.
We do not optimize $(x,y)$ directly; instead, a tiny neural network maps its parameters $p\in\mathbb{R}^3$ to an output pair $(x(p),y(p))$ by evaluating the MLP on two fixed inputs. 
Training therefore minimizes the composed loss $f(x(p),y(p))$, producing simultaneously a path in parameter space (left panel) and its image in output space (right panel). 
As discussed in the paragraph above, changes in the parameters are pushed forward to changes in $(x,y)$, while the loss information in output space is pulled back to guide the parameter update (here using a natural-gradient preconditioner). 
As a result, regions where $f$ has small gradients and saddle-like geometry in $\mathcal{Y}$ can appear as extended slow segments along the parameter trajectory, modulated by the local conditioning of $J(p)$. 
In the following section we will see how this toy example with $N=1$ mimics the way neural quantum states operate. 

\begin{figure*}[t]
  \centering
  \includegraphics[width=0.8\textwidth]{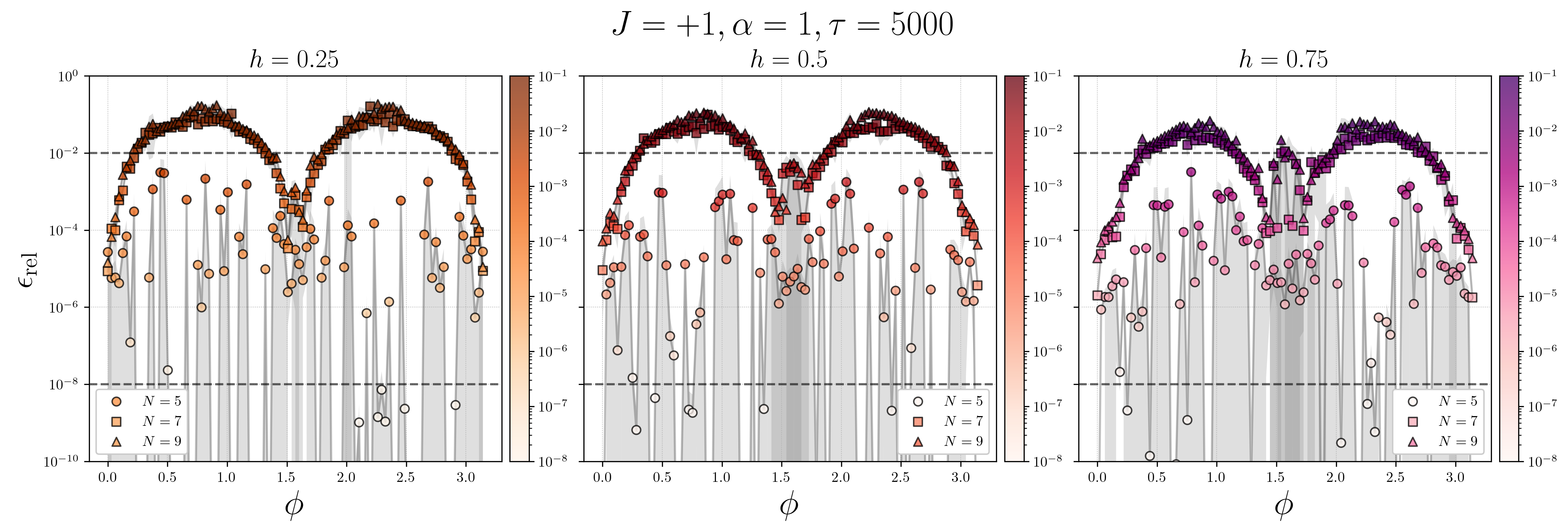}\par\smallskip
  \includegraphics[width=0.8\textwidth]{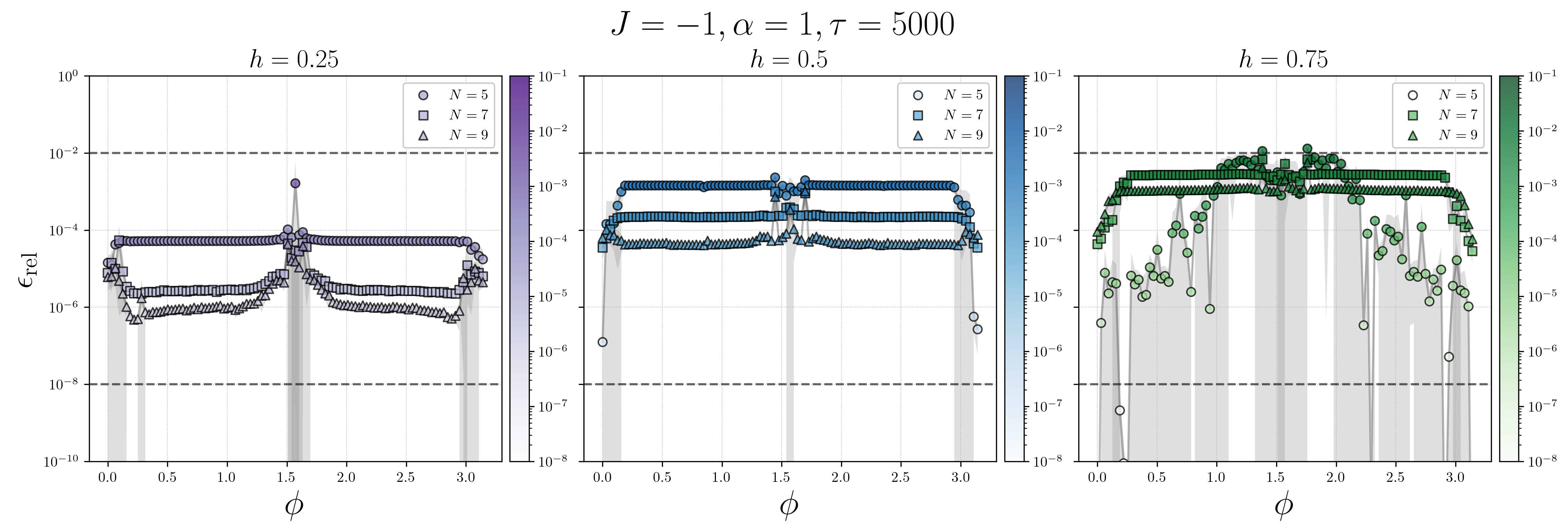}
\captionof{figure}{\justifying {\bf Energy minimization:} Relative error $\epsilon_{\mathrm{rel}}$ as a function of the rotation angle $\phi$ for an RBM ($\alpha=1$) optimized with stochastic reconfiguration (learning rate $\eta=10^{-2}$, regularization $\lambda=10^{-3}$) for $\tau=5000$ iterations using the energy objective for the rotated Hamiltonian $H(\phi)=U_\phi H U_\phi^\dagger$ in Eq.~\eqref{eq:Hamiltonian}. 
The optimization performance displays a strong dependence on the rotation angle $\phi$.
Upper panel: in the antiferromagnetic case $J=+1$, performance deteriorates for nearly all angles except the special points $\phi_k=0,\pi/2,\pi$, across different values of $h$ and system sizes $N$. 
Lower panel: in the ferromagnetic case $J=-1$, $\epsilon_{\mathrm{rel}}$ appears comparatively flat as a function of $\phi$ and decreases with increasing $N$. 
However, this behavior is misleading, since convergence toward the exact ground state is not achieved (see Sec.~\ref{sec:Results}).}
  \label{fig:rotation_results}
\end{figure*}

\subsection{Variational quantum models as smooth maps}\label{subsec:vq_smooth_maps}

Variational quantum ans\"atze, including neural quantum states (NQS)~\cite{Carleo_2017}, tensor networks, and parametrized quantum circuits (VQE)~\cite{Larocca_2023}, can all be described within the same geometric framework: a smooth parametrized map associating each parameter configuration with a many-body quantum state.
Concretely, for a parameter manifold $M\subset\mathbb{C}^P$ we consider
\begin{equation}\label{eq:vq_smooth_map}
    f: M \longrightarrow \mathcal{H}\simeq \mathbb{C}^{2^N}, \qquad \theta \longmapsto \ket{\psi(\theta)} ,
\end{equation}
where $\mathcal{H}$ is the $N$-qubit Hilbert space (with the understanding that physical states are defined up to normalization and global phase). 

As discussed, in the NQS case, the coordinate representation of $\ket{\psi(\theta)}$ in the computational basis is given by amplitudes $\psi_\theta(s)=\langle s\vert\psi(\theta)\rangle$ for configurations $s=(s_1,\dots,s_N)$ \cite{Carleo_2017}. 
In the VQE case, $f(\theta)$ is the state prepared by a parametrized circuit $U(\theta)$ acting on a reference state, $ \ket{\psi(\theta)}=U(\theta) \ket{0}^{\otimes N}$, while tensor-network ans\"atze define differentiable parametrizations of $\ket{\psi(\theta)}$ through their internal tensor structure.

This viewpoint provides a unified geometric description of optimization across NQS, tensor-network, and VQE approaches.
Although these variational families employ different parametrizations and optimization schemes, they can all be interpreted as inducing different geometric structures on parameter space through the pullback of the Fubini-Study metric.
Consequently, differences in trainability may emerge not only from expressive limitations of the ansatz, but also from the way each parametrization deforms the geometry of the underlying variational manifold.
\begin{figure*}[t]
  \centering
  \includegraphics[width=0.8\textwidth]{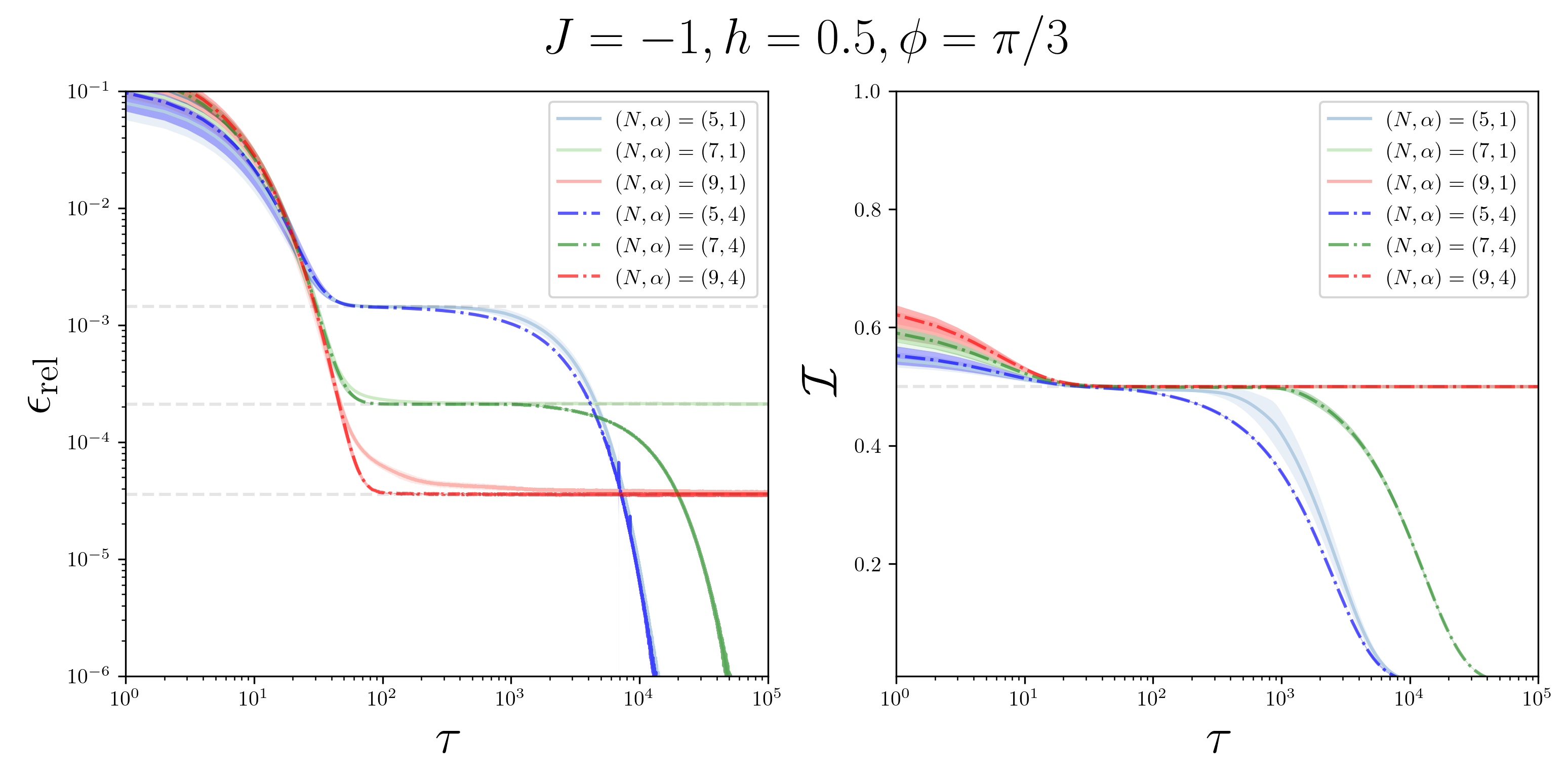}
\captionof{figure}{\justifying Left panel: log-log plot of the averaged relative energy error over $10$ samples for $N=5,7,9$ using an RBM architecture ($\alpha=4$), optimized for $\tau=10^5$ stochastic reconfiguration iterations with $\eta=10^{-2}$ and $\delta = 10^{-6}$. 
The horizontal dashed lines indicate the relative energy difference between the exact ground state and the first excited state divided by two, corresponding to the saddle-point region reminiscent of the toy example in Fig.~\ref{fig:toy_example}.
Right panel: the infidelity $\mathcal{I}$ with respect to the exact ground state remains trapped near $0.5$. 
Increasing the parameter ratio $\alpha_{\mathrm{RBM}}$ partially alleviates the trapping, but the number of optimization steps required for convergence rapidly becomes prohibitive as the system size increases.}
  \label{fig:saddle_point}
\end{figure*}

\subsection{Numerical Setup and Validation}

Throughout this work we focus on the solvable transverse-field Ising Hamiltonian in Eq.~\eqref{eq:Hamiltonian}, for which both the spectral properties and the basis dependence induced by the rotation angle $\phi$ can be controlled analytically.
Nevertheless there are multiple numerical considerations that need to be taken into account.
\begin{itemize}
    \item \textit{Sampling Noise} - Although Monte Carlo sampling is common in large‐scale NQS, here we avoid it by restricting to \(N<20\), ensuring exact gradient and expectation evaluation.
    \item \textit{Architecture Limitations} - Shallow RBMs and small FCNNs may lack the expressivity required to represent complex rotated states. Tuning the hidden‐unit ratio (\(\alpha_{\text{RBM}}=M/N\)) or network depth is crucial, but even well‐tuned models can stall when the target lies in a geometrically unfavorable region.
    \item \textit{Quantum Natural Gradient Conditioning} - The quantum Fisher matrix $G(\alpha)$ can become ill‐conditioned in narrow valleys or near saddles, causing updates to stall or oscillate. Regularization ($\epsilon I$) mitigates this but does not fully eliminate the issue.
    \item \textit{Initialization Bias} - Neural network parameters are typically initialized randomly near zero, which yields an initial state close to the equal‐weight superposition 
    \begin{equation}
        \lvert W\rangle
  = \frac{1}{\sqrt{2^{N}}}\sum_{s\in\{\pm1\}^N} \lvert s\rangle.
    \label{WDef}
    \end{equation}
To remove dependency on initialization, we first pretrain the chosen NQS models (in this case RBM and a shallow FCNN) to represent $\lvert W\rangle$ by minimizing its infidelity. Once convergence is achieved (infidelity $<10^{-8}$), these parameters serve as a common, exact initialization for all subsequent runs. Without pretraining, different random seeds can yield diverse trajectories. By pretraining to $\lvert W\rangle$, we fix a deterministic initialization, isolating the effect of rotated targets on convergence.
\end{itemize}
By controlling these numerical effects, we ensure that the observed differences in training performance can be directly attributed to the geometry of the variational landscape and to the relative location of the rotated target states within the variational manifold, rather than to extraneous numerical artifacts.

\begin{figure*}[t]
  \includegraphics[width=0.75\textwidth]{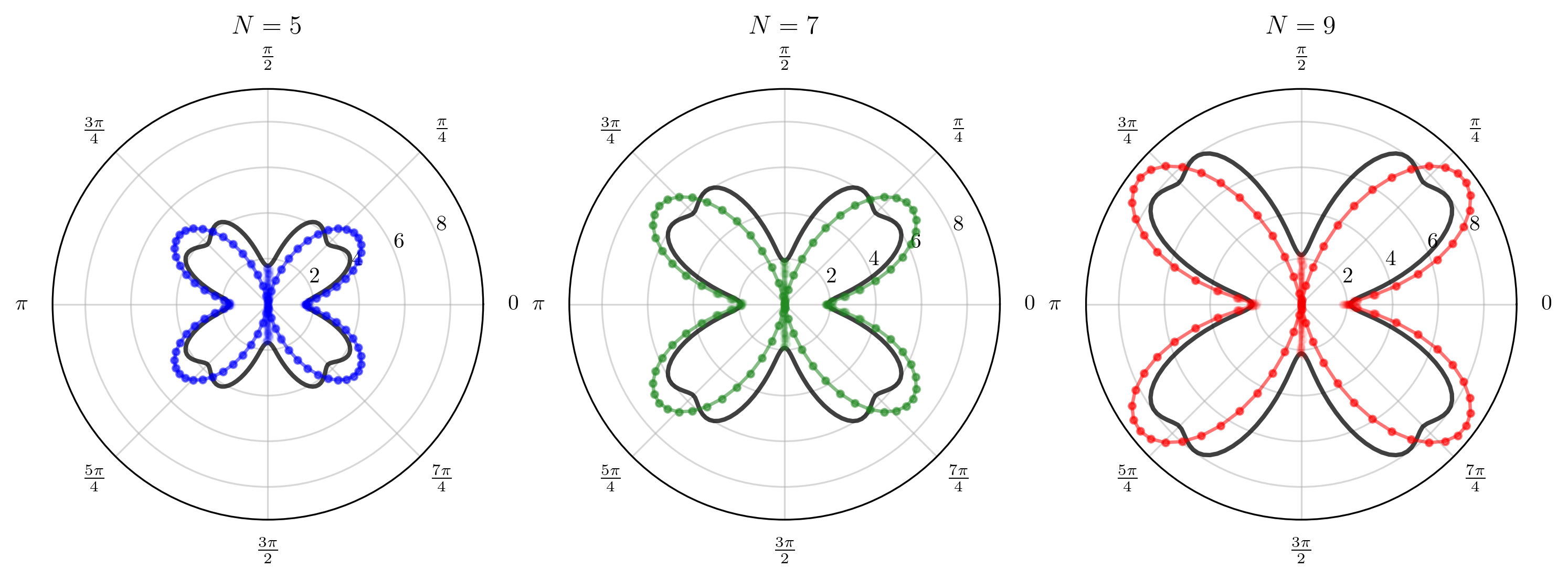}\par\smallskip
  \includegraphics[width=0.75\textwidth]{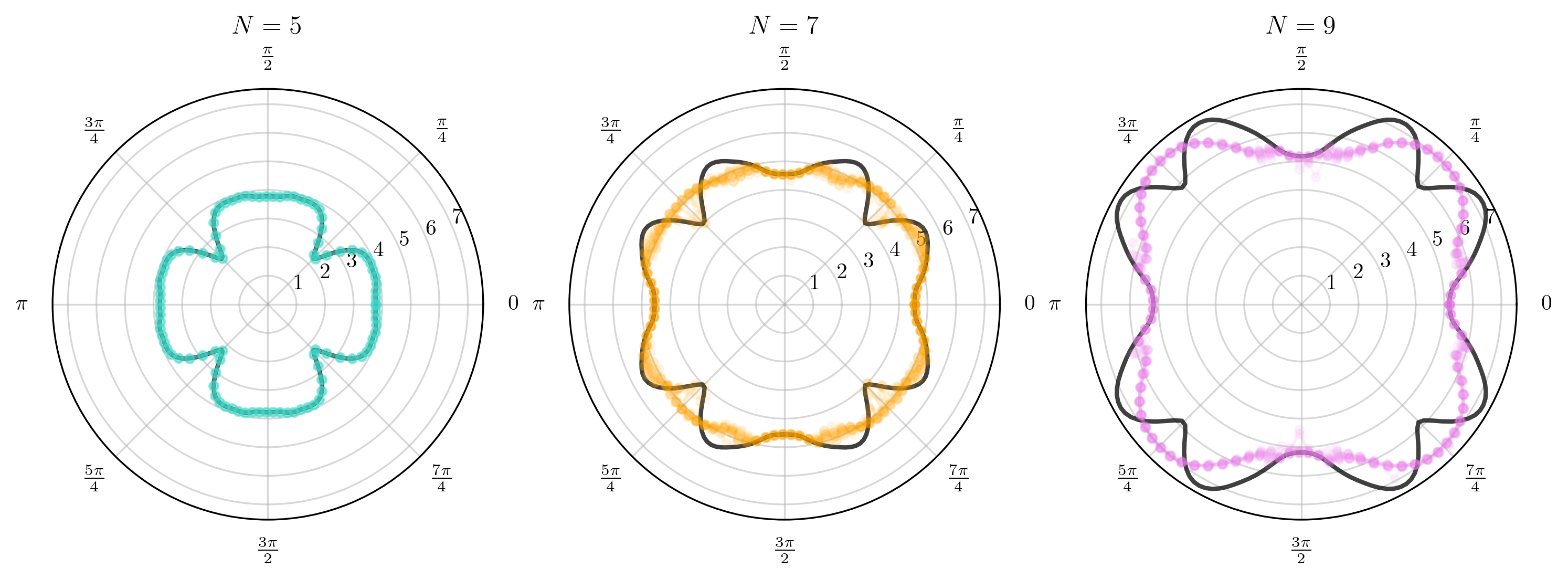}
\captionof{figure}{\justifying Shannon entropy of the variational state-vector coefficients (quantum coherence) for an RBM ($\alpha=1$) optimized with stochastic reconfiguration using energy minimization for $\tau=5000$ iterations, with $h=0.5$, $N=5,7,9$, and couplings $J=-1$ (top panels) and $J=+1$ (bottom panels), as a function of the rotation angle $\phi$. 
In the ferromagnetic case ($J=-1$), the entropy of the variational states converging toward superpositions of the ground and first excited states strongly deviates from the exact ground-state value (black curve).
A similar deviation progressively emerges for larger system sizes also in the antiferromagnetic case ($J=+1$).}
  \label{fig:coherence}
\end{figure*}

\section{Results}\label{sec:Results}

We start by directly comparing the information‐geometric distance introduced in Sec.~\ref{sec:considerations} between the selected initialization state $\psi\equiv\lvert W\rangle $ from Eq.~\eqref{WDef} and the target state $\varphi(\phi)\equiv\lvert \psi_{\phi}\rangle$, using the infidelity cost function in Eq.~\eqref{eq:infidelity}. 
By using a fixed initial state and the corresponding parametrization, we can clearly show how rotating the target state $\varphi$ changes the distance between the two states. 
Results are shown in Fig.~\ref{fig:FS_comparison}, where we quantify how basis rotation affects the optimization trajectory. 
Changing the interaction parameter $J=\pm1$, we can compare how the path from the same initial state depends on the distance to differently rotated target. Comparing RBM and a shallow FFNN architecture we find similar results. 

We analyze the infidelity cost function (denoted by $\mathcal{L}$) both as a function of the optimization steps $\tau$ and of the rotation angle $\phi$. 
In the upper-left panel of Fig.~\ref{fig:FS_comparison} ($J=-1$), the trajectories corresponding to angles $\phi=0,\pi$ remain close to the orthogonal value $\mathcal{L}=1$. 
The right panels show that the Fubini--Study distance $\gamma(\psi,\varphi)$~\cite{Stokes_2020} becomes maximal in the vicinity of these angles, indicating that the optimization slowdown is directly correlated with the geometric separation between the initialization and target states. 
Switching to the antiferromagnetic case $J=+1$ further strengthens this correspondence, since the range of angles characterized by large geometric distance closely matches the region where optimization becomes trapped.

While infidelity minimization provides a direct probe of trainability~\cite{Sinibaldi_2023}, in practical applications the target state is typically unknown and optimization must instead rely on the variational energy in Eq.~\eqref{eq:energy}. 
As already illustrated in Fig.~\ref{fig:umap_trajectories}, different objective functions generate substantially different optimization landscapes. 
Consequently, the difficult regions identified through infidelity minimization do not necessarily coincide with those encountered during energy optimization.

Comparing the cases $J = \pm 1$ also allows us to study two models with the same underlying symmetries but qualitatively different low-energy spectra: the gapped ferromagnetic case ($J = -1$) and the gapless antiferromagnetic case ($J = +1$). 
In the ferromagnetic regime, the energy separation between the two lowest eigenstates decreases exponentially with system size $N$, whereas in the antiferromagnetic regime the low-energy band closes only algebraically.

In Fig.~\ref{fig:rotation_results} we use the relative energy error
\[
\epsilon_{\mathrm{rel}}
=
\left|
\frac{E_{\mathrm{NQS}}-E_{\mathrm{exact}}}{E_{\mathrm{exact}}}
\right|
\]
as the main performance indicator.

Let us first consider the antiferromagnetic case (upper panels of Fig.~\ref{fig:rotation_results}). 
For larger system sizes $L>5$, the RBM fails to converge for nearly all rotation angles except the special points $\phi_k=0,\pi/2,\pi$. 
In these regions, the shallow underparametrized NQS is unable to accurately represent the rotated ground state within the allotted number of optimization steps.

On the other hand, the ferromagnetic case (lower panels) appears to display higher accuracy together with a comparatively weak dependence on the rotation angle $\phi$, while the relative energy error seemingly decreases with increasing system size. 
However, this behavior is deceptive.
Comparing both the relative energy error and the infidelity for angles within this apparently stable regime, for example at $\phi=\pi/3$ in Fig.~\ref{fig:saddle_point}, it becomes evident that the ansatz has not converged to the true ground state, since the infidelity $1-\mathcal{F}$ remains $\gtrsim0.5$. 

Instead, the variational state converges toward a superposition of the two lowest-energy eigenstates, whose energy separation decreases exponentially with system size.
This observation highlights how, in the present setting, convergence properties depend strongly on the low-energy structure of the Hamiltonian and on the local geometric features of the variational landscape.
In particular, near-degenerate regions generate low-curvature sectors and saddle-like structures that are difficult to escape from~\cite{Dauphin_2014,McClean_2018}.
The proposed rotation of the target state further increases the probability that the optimization becomes trapped in these regions. 
While increasing the parameter ratio $\alpha_{\mathrm{RBM}}$ partially alleviates the trapping, the number of optimization steps required for convergence rapidly becomes impractical even for relatively small systems such as $L=9$ when using shallow architectures.
The problem that emerges in this context is the quality of the obtained state. 
Rotating the basis does not modify entanglement measures that depend only on the spectrum of reduced density matrices; for instance, the entanglement entropy of any bipartition is identical for $\lvert \psi_{0}\rangle$ and $\lvert \psi_{\phi}\rangle$.  
However, other global quantum resources, such as non-stabilizerness and quantum coherence~\cite{Kozic_2025,Kozic_2026}, may vary substantially depending on whether one considers the exact ground state or an arbitrary superposition of nearby eigenstates.

To quantify this effect, we use quantum coherence, defined as the Shannon entropy of the wavefunction coefficients in the rotated basis, as an additional diagnostic tool.
Since $U_{y}(\phi)$ mixes $\lvert 0\rangle$ and $\lvert 1\rangle$ locally at each site, the rotated ground state generally acquires a nontrivial amplitude distribution and its coherence may significantly increase relative to the unrotated case. 
In Fig.~\ref{fig:coherence} we show that the quantum coherence of the variational states obtained after a fixed number of optimization steps can differ substantially from the exact value, even for very small systems.

\section{Conclusion and Discussion}\label{sec:Conclusion}

 \begin{figure*}[t]
    \centering
  \includegraphics[width=0.9\linewidth]{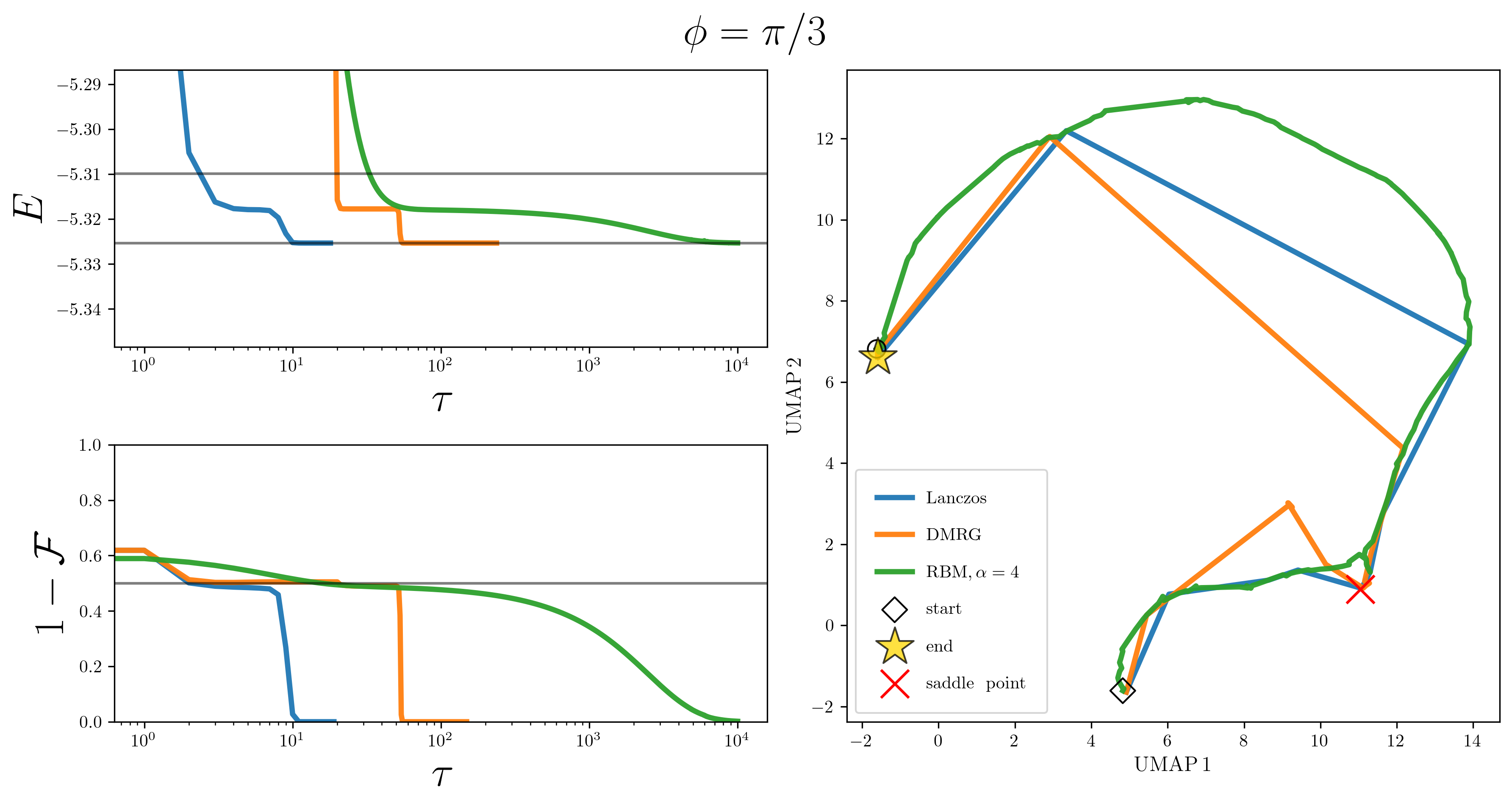}
  \caption{\justifying Discussion in Sec. \ref{subsec:vq_smooth_maps} can help us unify the understanding in variational optimization of different algorithms. In this case we compare the performance of Lanczos algorithm~\cite{Lanczos_1950}, DMRG~\cite{White_1992} and RBM with $\alpha=4$ starting from the same initial statevector and the ferromagnetic Ising model $N=5, h=-0.5, \phi=\pi/3$. Top left panel shows energy $E$ during iterations $\tau$ while bottom plot shows infidelity $\mathcal{I}$. All three algorithms encounter the same saddle point region. The right plot compares the corresponding statevector trajectories using UMAP projection.}
  \label{fig:umap_algorithms}
\end{figure*}

The importance of the proliferation of saddle points and the connection to computational complexity extends beyond NQS and can also be observed in the comparison between different variational algorithms (see Fig.~\ref{fig:umap_algorithms}). 
By treating different variational ans\"atze within the common geometric framework discussed in Sec.~\ref{subsec:vq_smooth_maps}, we initialize state-vector methods, tensor networks, and NQS from the same initial state. 
Although these approaches rely on different parametrizations and optimization schemes~\cite{Larocca_2023}, all of them encounter qualitatively similar saddle-point regions, suggesting that optimization slowdowns may arise from analogous geometric obstructions despite the different parametrizations and update rules employed by the algorithms.

In this work, we investigated how local basis rotations in a one-dimensional transverse Ising model affect the convergence of Neural Quantum States without modifying the underlying energy landscape.
By applying sitewise rotations parameterized by an angle $\theta$, we displaced the exact ground-state wavefunction within a fixed optimization landscape and analyzed the response of shallow variational ans\"atze, in particular Restricted Boltzmann Machines and small feedforward networks trained with quantum natural gradient.
This construction allowed us to separate optimization-induced effects from changes in the physical properties of the target state.
Although the energy spectrum remains unchanged, the information-geometric distance between the rotated wavefunction and typical initial variational states increases, making convergence toward the exact ground state progressively more challenging.

In the ferromagnetic regime, near-degenerate low-energy eigenstates generate poorly conditioned regions and high-curvature barriers in parameter space that manifest as saddle points.
As a consequence, the optimization frequently converges to low-energy superpositions that fail to reproduce the correct coefficient statistics, despite exhibiting small relative energy errors.
This mismatch highlights the limitations of energy-based metrics in diagnosing the quality of the learned wavefunction.
Similar geometric obstructions persist across rotation angles and in both ferromagnetic and antiferromagnetic regimes, highlighting the strong sensitivity of NQS optimization to the location of the target state within parameter space rather than to changes in the objective function itself.

Several practical implications follow from these findings.
First, basis rotations that leave the energy landscape invariant can nevertheless degrade variational performance by relocating the target state into regions characterized by unfavorable curvature, where quantum natural gradient updates become ineffective.
Second, shallow architectures that perform reliably near the computational basis show pronounced degradation when required to represent rotated states with enhanced quantum resources, such as coherence or non-stabilizerness.
Third, optimizing infidelity rather than energy improves sensitivity to the structure of the target state but does not fully suppress trapping near saddle points induced by geometric displacement.
Taken together, these observations indicate that trainability is strongly influenced by the geometric relation between the target state and the variational manifold.

Looking forward, our results suggest several strategies for improving variational optimization.
More expressive architectures, such as deeper feedforward networks or convolutional structures, may better navigate high-curvature regions and accurately represent rotated states carrying larger quantum resources.
In addition, adaptive regularization of quantum natural gradient updates or hybrid objective functions combining energy and infidelity could help mitigate convergence toward suboptimal basins.
Finally, analytically solvable models such as the rotated Ising chain provide a valuable testbed for diagnosing optimization failures, enabling systematic studies of variational landscape geometry and guiding the development of robust and interpretable variational algorithms for complex quantum systems.

More generally, our results identify a geometric contribution to basis dependence in Neural Quantum States, beyond limitations purely associated with representational expressivity. 
In this perspective, controlled basis rotations provide a simple and analytically transparent probe of trainability geometry in variational quantum many-body methods.

\section*{Acknowledgement}
SBK thanks Gianpaolo Torre and Jovan Odavić for their help and support, and also thanks Miha Srdinšek, Sidhartha Shankar Dash, Giuseppe Carleo and Roberto Verdel for insightful discussions and the support from the Croatian Science Foundation Projects No. DOK-2020-01-9938. 
VZ acknowledges support from the Croatian Science Foundation project No. IP-2022-10-1648: "Stochastic processes on networks analysis in systems with limited information".
FF and SMG acknowledge support from the Croatian Science Foundation through the project No. IP-2025-02-1667 titled:"Mining the Quantum: Frustration, Disorder, and Devices".
VZ, FF, and SMG also acknowledge support from the project "Implementation of cutting-edge research and its application as part of the Scientific Center of Excellence for Quantum and Complex Systems, and Representations of Lie Algebras", Grant No. PK.1.1.10.0004, co-financed by the European Union through the European Regional Development Fund - Competitiveness and Cohesion Programme 2021-2027.

\section*{Data and Code Availability}
Data and codes are available upon request to the authors or at the Zenodo repository~\cite{dataset}. Some of the simulations were partially performed using the open source Netket library~\cite{Vincentini_2022_01, Vincentini_2022_02}, while majority used custom adapted codes in Jax framework~\cite{Jax_2018}.

\bibliographystyle{unsrt}
\bibliography{refs}

\end{document}